\documentclass[12pt,a4paper]{iopart}
\usepackage{amssymb,graphicx,cite}
\eqnobysec
\begin{document}


 
 


 
 
   
 

\title[Collapse in a fermion-fermion mixture]{Dynamical collapse 
in  a degenerate binary fermion 
mixture using a hydrodynamic model}
 
\author{Sadhan K. Adhikari\footnote{Electronic
address: adhikari@ift.unesp.br; \\
URL: http://www.ift.unesp.br/users/adhikari/}}
\address
{Instituto de F\'{\i}sica Te\'orica, UNESP $-$ S\~ao Paulo State
University,
01.405-900 S\~ao Paulo, S\~ao Paulo, Brazil\\}

\date{\today}
 
 
\begin{abstract}

We  use a time-dependent dynamical hydrodynamic
model to study
a collapse  in a degenerate fermion-fermion       
mixture (DFFM) of different atoms. Due to a strong Pauli-blocking
repulsion among
identical spin-polarized fermions at short distances there cannot be
a collapse  for  repulsive interspecies
fermion-fermion interaction.
However, 
there can be
a collapse  for a sufficiently attractive  interspecies
fermion-fermion interaction in a DFFM  of different 
atoms. Using a variational analysis and numerical solution of the 
hydrodynamic model we study
different aspects of   collapse in such a DFFM
initiated by a jump in the interspecies
fermion-fermion interaction (scattering length)
to a large negative (attractive) value  using a
Feshbach resonance. Suggestion for experiments of collapse in a DFFM
 of distinct atoms 
is
made.

\end{abstract}


\maketitle

 \section{Introduction}
 
Recent successful observation of degenerate  boson-fermion  mixture 
(DBFM) and 
fermion-fermion  mixture (DFFM) of
trapped alkali-metal atoms by different experimental groups
\cite{exp1,exp2,exp21,exp211,exp22,exp3,exp4} has initiated the 
intensive 
experimental
studies of different novel phenomena \cite{exp6,exp5,exp5x}.  
It has been possible to achieve a degenerate Fermi gas  (DFG) by 
sympathetic cooling in the presence of a second boson or fermion 
component, as there cannot be an effective evaporative cooling \cite{exp1} 
of a 
single-component DFG  due to a strong Pauli-blocking
repulsion at low 
temperature  among spin-polarized fermions.  
Among these
experiments on a DFG, apart from the study of a DBFM in
$^{6,7}$Li \cite{exp3}, $^{23}$Na-$^6$Li
\cite{exp4} and $^{87}$Rb-$^{40}$K 
\cite{exp5,exp5x,bongs,mur},  
there have been studies of a DFFM 
in
$^{40}$K-$^{40}$K$^*$ \cite{exp1} and $^6$Li-$^6$Li$^*$ 
\cite{exp2,exp21,exp211,exp22} 
systems, where $^*$ denotes a distinct  hyperfine state. More 
recently the
formation
of a Bardeen-Cooper-Schreiffer (BCS) condensate of fermionic $^6$Li
atoms in  $^{23}$Na-$^6$Li \cite{bcs}
 and $^6$Li-$^6$Li$^*$ \cite{exp211}
mixtures 
has been observed experimentally.
 The collapse in a DBFM  of 
$^{87}$Rb-$^{40}$K atoms has been observed and studied by Modugno {\it et
al.}
\cite{exp5} and more recently by Ospelkaus  {\it et
al.}
\cite{bongs}. 
Recently, experiments on controlled collapse on $^{87}$Rb-$^{40}$K
have been accomplished \cite{ccol}.
 
A collapse in a Bose-Einstein condensate
(BEC) takes place
due to an attractive atomic
interaction \cite{hulet,don}.  
A study of controlled collapse   and explosion
has been performed by Donley {\it et al.}
\cite{don} on an attractive  $^{85}$Rb BEC, where they
manipulated the inter-atomic interaction by varying
a background magnetic
field exploiting  a nearby Feshbach resonance 
\cite{fs}. 
There
have been many  theoretical \cite{th1,th2} studies to describe different
features of this experiment \cite{don}. 
More recently, there have been experimental studies on collapse in a DBFM
of $^{87}$Rb-$^{40}$K by two different groups
\cite{exp5,bongs,ccol} as 
well as related theoretical investigations \cite{ska,zzz}.
As the interaction in a pure DFG  at 
short
distances is
repulsive due to Pauli-blocking, there cannot be a
collapse in
it. The Pauli repulsion is responsible for the stability of a
neutron star
against a (gravitational) collapse. 
A collapse is possible  in a DBFM
in the presence of a sufficiently strong
boson-fermion attraction which can overcome 
the Pauli repulsion among
identical fermions \cite{exp5,bongs,ska}.

In this paper we study the  collapse in a DFFM
for a sufficiently attractive interspecies fermion-fermion interaction
which can overcome the Pauli repulsion.  However, there is already 
experimental
evidence and theoretical conjecture that a Fermi gas in two (spin) 
hyperfine
states of the same atom is much more stable \cite{gehm} than expected on
the basis of a scattering length approach \cite{exp22,phase} and there
is no collapse in such a system. A similar conclusion follows from an
examination of the compressibility of such a system \cite{kok}.  A
strongly attractive DFFM in two (spin)
hyperfine
states
exhibits universal behavior and should be mechanically stable as a
consequence of the quantum-mechanical requirement of unitarity. This
requirement limits the maximum attractive force for such a  DFFM
to a value smaller than that of the outward Fermi pressure due to Pauli
repulsion \cite{gehm}.  It has been demonstrated that a two-component
DFFM in different (spin) hyperfine states is stable against collapse
\cite{hei}, whereas a multicomponent degenerate fermion mixture
\cite{gehm,hei} or a DFFM of different atoms \cite{hulet1} could undergo
collapse. Hence, by taking a DFFM of two different fermionic atoms of
different atomic mass one can avoid the
problem \cite{gehm,hei} of a possible suppression of collapse. Thus one
could study collapse in a DFFM in the same manner as in a DBFM. The
second component of fermions will
then only aid in inducing an attraction among the fermions of the first
component responsible for collapse without suppressing the collapse.
Although, the past experiment \cite{gehm} on two-component cold Fermi 
gas
used two (spin) hyperfine states of the same atom, experiments can be 
realized
with distinct atoms and one can look for collapse in such a system. One 
such system is the $^6$Li-$^{40}$K mixture: both $^6$Li \cite{exp211} 
and 
$^{40}$K \cite{exp1}
have been trapped and studied in laboratory.

Here we use a coupled time-dependent mean-field hydrodynamic model which 
is
inspired by the success of a similar model used by the present author in
the investigation of a fermionic collapse \cite{ska} and bright
\cite{fbs2} and dark \cite{fds} solitons in a DBFM as well as of
mixing-demixing \cite{md} and black solitons \cite{bs} in a DFFM. The
conclusions of the study on bright soliton \cite{fbs2} are in agreement
with a microscopic study \cite{bong1} and the noted survival of collapse
in the numerical study \cite{ska} has been experimentally substantiated
later in a DBFM of $^{40}$K-$^{87}$Rb \cite{bongs}.  A very similar
model has been used by Jezek {\it et al.} \cite{Jezek} in a successful
description of vortex states in a DBFM. Although, a mean-field model is 
simple to use and leads to a proper prediction of probability density of 
the fermionic system, many true quantum effects are lost in this simplified 
model, e. g., it cannot predict the suppression of collapse of a DFFM in 
two different (spin) hyperfine states \cite{gehm,hei} as discussed in 
the last 
paragraph. We recall  many true 
quantum effects are also 
lost \cite{11}  in the mean-field Gross-Pitaevskii equation for trapped 
bosons.

In
our study on collapse in a DFFM we shall consider a strong attraction
among fermions which will naturally lead to molecule formation and not
to a BCS state. A BCS state is usually formed for a weak attraction
among identical fermions. The possibility of molecule formation by 
three-body recombination 
is  explicitly included in our model by an absorptive nonlinear term. 
Apart from the direct experimental interest in
trapped cold atoms, a study of strongly interacting Fermi gases and
their possible collapse is also relevant \cite{gehm} in condensed matter
physics (superconductivity), nuclear physics (nuclear matter), high
energy physics (effective theories of strong interaction), and
astrophysics (compact stellar objects), which makes the present study of
greater interest.

The collapse in a  DFFM
is first studied  by a 
variational analysis of the present model, which is later substantiated by 
a complete numerical solution using the Crank-Nicholson scheme \cite{sk1}. 
During a collapse and an explosion of the  
DFFM, the system loses atoms as 
in a collapsing and exploding BEC \cite{don}. The loss of atoms is 
accounted for by three-body 
recombination involving two types of fermions. We study the 
sensitivity 
of our results on the three-body recombination
loss rates. We also study the quasi-periodic 
oscillation of the sizes of the  
DFFM
undergoing a
collapse. The collapse 
is to be
initiated by jumping the interspecies scattering length to a large 
negative (attractive) value near a fermion-fermion Feshbach resonance 
\cite{fsff}. However, the collapse starts after a time delay upon this 
jump and we study the variation of this time delay with the final 
scattering length. This variation has a behavior similar to that observed 
in the bosonic case \cite{don}. The collapsing  
DFFM 
is found to execute a 
quasi-periodic oscillation with a frequency approximately equal to twice 
the frequency 
of the harmonic trap as in a BEC \cite{don}.

Previously, in addition to the study of a collapse in a pure 
BEC \cite{th1,th2},
we also investigated \cite{adhi} the collapse   in a
two-species
BEC initiated by 
an interspecies attraction. 
The predicted collapse in a two-species  BEC for intra-species
repulsion  and interspecies attraction \cite{adhi}
is similar to that in a  DBFM
of $^{87}$Rb-$^{40}$K
studied before 
\cite{exp5,bongs,ska} and 
that in a  DFFM
studied in
this paper. 

In section 2 we present the coupled hydrodynamic model for a  DFFM
 which we 
apply to predict and study a collapse. In this
section we also present a variational analysis based on this model
which substantiates the
collapse in a  DFFM
for a sufficiently attractive
interspecies
fermion-fermion attraction.  In section 3 
we 
present results of numerical simulation of  our study on collapse. 
Finally, in section 4  we present a brief summary  of our 
investigation. 

 \section{Coupled Hydrodynamic Model for a Fermion-Fermion Mixture}

\subsection{Model}

A mean-field-hydrodynamic Lagrangian for
a DFG   has been used successfully in the study of a
DBFM
\cite{fbs2,fds,Jezek}
which we shall use in the present investigation. The virtue of the
hydrodynamic  model for a DFG  over a  microscopic
description is its 
simplicity and
good predictive power.
To  develop a set of practical time-dependent hydrodynamic
equations for a  DFFM, we consider   the
following Lagrangian density \cite{fbs2,fds}
\begin{eqnarray}\label{yy}
&{\cal L}& = \frac{i}{2}\hbar \sum_{j=1,2}\left(
\Psi_j \frac{\partial \Psi_j \- ^*}{\partial t} - \Psi_j \- ^  
*
\frac{\partial \Psi_j
}{\partial
t} 
\right) \nonumber \\ 
&+& \sum_{j=1}^2 \left(\frac{\hbar^2 |\nabla
  \Psi_j |^2 }{6m_j}+
V_j n_j+\frac{3}{5}  A_j  |n_j |^{5/3}\right)\nonumber \\
&+& g_{12}  n_1 n_2-i\hbar \left(K_{31}n_1 n_2^2  
+K_{32} n_1^2 n_2  \right),  
\end{eqnarray}
where  $m_j$ is the
mass of component $j (=1,2)$, 
$A_j=\hbar^2(6\pi^2)^{2/3}/(2m_j)$, 
$\Psi_j$ is a complex probability amplitude,  $n_j=|\Psi_j|^2$ is a
real probability density  and
$N_j \equiv  \int d{\bf r} n_j({\bf r}) $  the number.  
Here the  interspecies
coupling is 
$g_{12}=2\pi \hbar^2 a_{12}
/m_R$
with the
reduced mass $m_R=m_1m_2/(m_1+m_2),$ and  $ a_{12}$
is the interspecies
scattering length.
The spherically-symmetric potential is taken as  $
V_{j}({\bf
r})=\frac{1}{2}(3m_j) \omega ^2 r^2$ where
 $\omega$ is  the radial ($r$) frequency.
The interaction between intra-species fermions in
spin-polarized state is highly suppressed due
to Pauli blocking
and has been neglected in  (\ref{yy})  and will be
neglected throughout.
The kinetic energy terms in this equation
are derived from a hydrodynamic equation for the
fermions
\cite{capu}. The
kinetic energy terms contribute little to this problem compared to the
dominating Pauli blocking term 
$3A_j|n_j|^{5/3}/5$  in (\ref{yy}).
However, its inclusion leads
to a smooth  solution for the probability density everywhere
\cite{fbs2}. The Lagrangian density of each fermion component in
 (\ref{yy}) is
identical to that used in Refs.  \cite{fbs2,fds}. The last two terms in
 (\ref{yy}) correspond to three-body recombination  due to the
following reactions, respectively
$F_1+F_2+F_2  \to  (F_1F_2)+F_2,$ and 
$F_1+F_2+F_1  \to  (F_1F_2)+F_1,$
where $(F_1F_2)$ is a composite structure (resonance/molecule) of
fermions $F_1$ and $F_2$ and $K_{31}$ and $K_{32}$ are the corresponding
three-body
loss rates.  The contribution to the Lagrangian density of the
recombination reactions 
is proportional to
the density of the participating fermions. Here we neglected
two-body loss.

The  dynamical equations for the system are just the 
usual  Euler-Lagrange  (EL) equations with the Lagrangian density 
(\ref{yy}) \cite{gold}
\begin{equation}
\frac{\partial}{\partial t}\frac{\partial {\cal L}}{\partial
\frac{\partial
\Psi_j\- ^*}{\partial t}}+
\sum _{k=1}^3 \frac{d}{dx_k}\frac{\partial {\cal L}}{\partial
\frac{\partial \Psi_j\- ^*}{\partial x_k}}= \frac{\partial {\cal
L}}{\partial
 \Psi_j\- ^*},
\end{equation}
where $x_k, k=1,2,3$ are the three space components, and 
$j=1,2$ refer to the fermion components. 
Consequently,   the following  EL   equations of 
motion are
derived:
\begin{eqnarray}\label{e} \biggr[ &-& i\hbar\frac{\partial
}{\partial t}
-\frac{\hbar^2\nabla_{\bf r}^2}{6m_{{1}}}
+ V_{{1}}({\bf r})+A_1|n_1|^{2/3} 
+ g_{{12}} n_2
\nonumber \\
&-&i\hbar \left( K_{31}n_2^2+2K_{32}n_1  n_2 \right)
 \biggr]\Psi_1=0. 
\end{eqnarray}
\begin{eqnarray}\label{f} \biggr[& -& i\hbar\frac{\partial
}{\partial t}
-\frac{\hbar^2\nabla_{\bf r}^2}{6m_{{2}}}
+ V_{{2}}({\bf r}) 
+ A_2 |n_2|^{2/3} 
+ g_{{12}} n_1
\nonumber \\
&-& i\hbar \left( 2K_{31}n_1 n_2 + K_{32} n_1 ^2  \right)
 \biggr]\Psi_2 =0. 
\end{eqnarray}

In the spherically-symmetric state the fermion density  
 has the form $\Psi_j({\bf
r};t)=\psi_j(r;t).$ 
Now  transforming to
dimensionless variables
defined by $x =\sqrt 2 r/l$,     $\tau=t \omega, $
$l\equiv \sqrt {\hbar/(m\omega)}$, $m=3m_1=3m_2$
and
\begin{equation}\label{wf}
\frac{ \phi_j(x;\tau)}{x} =  
\sqrt{\frac{4 \pi l^3}{N_j\sqrt 8}}\psi_j(r;t),
\end{equation}
we obtain from  (\ref{e}) and (\ref{f})
\begin{eqnarray}\label{d1}
&\biggr[&-i\frac{\partial
}{\partial \tau} -\frac{\partial^2}{\partial
x^2} 
+\frac{x^2}{4} 
+ {\cal N}_{11}
\left|\frac {\phi_1}{x}\right|^{4/3}
\nonumber \\          
&+& 6 \sqrt 2    {\cal N}_{12}\left
|\frac {\phi_2}{x}\right|^2 
- i2\xi_{32}  {\cal N}_{12}  {\cal
N}_{21} \left|\frac
{\phi_1}{x}\right|^2  \left|\frac
{\phi_2}{x}\right|^2
 \nonumber \\  &-& 
i\xi_{31} {\cal N}_{12} ^2  \left|\frac
{\phi_2}{x}\right|^4
\biggr]\phi_1({ x};\tau)=0,
\end{eqnarray}
\begin{eqnarray}\label{d2}
&\biggr[&-i\frac{\partial
}{\partial \tau} -\frac{\partial^2}{\partial
x^2} 
+\frac{x^2}{4} 
+ {\cal N}_{22}
\left|\frac {\phi_2}{x}\right|^{4/3}
\nonumber \\   
&+& 6 \sqrt 2    {\cal N}_{21}\left
|\frac {\phi_1}{x}\right|^2
- i2\xi_{31}  {\cal N}_{12}  {\cal
N}_{21} \left|\frac
{\phi_1}{x}\right|^2  \left|\frac
{\phi_2}{x}\right|^2
 \nonumber \\  &-&
i\xi_{32} {\cal N}_{21} ^2  \left|\frac
{\phi_1}{x}\right|^4    
 \biggr] \phi_2({ x};\tau)=0,
\end{eqnarray}
where
$ {\cal N}_{jj}=3(3\pi N_j/2)^{2/3}$, 
$ {\cal N}_{12} =   N_2 a_{12} /l,$ 
$ {\cal N}_{21} =   N_1 a_{12} /l,$
$\xi_{32}=K_{32}/(2\pi^2 a_{12}^2l^4\omega),$ and
$\xi_{31}=K_{31}/(2\pi^2 a_{12}^2l^4\omega).$ In the non-absorptive 
case without any loss of atoms due to three-body recombination
$\xi_{31}=\xi_{32}=0$, the normalization of the wave-function components
is
given by $\int_0^\infty dx |\phi_j(x;\tau)|^2 =1, j=1,2.$ In the
absorptive case $\xi_{31}\ne 0$ and $\xi_{32} \ne 0$ it is possible to
have loss  of atoms due to three-body recombination and  the
normalization reduces with time due to loss of atoms. 

We solve the coupled hydrodynamic equations (\ref{d1}) and (\ref{d2})
numerically using a time-iteration method based on the Crank-Nicholson
discretization scheme elaborated in Refs. \cite{sk1,sk2}.  We discretize
the
hydrodynamic equations using time step $0.00025$ and space step $0.025$
 spanning $x$ from 0 to 25. This domain of space was sufficient to
encompass the entire fermion function during a collapse and explosion
and obtain convergent solution for the total number of fermions.  First
we solve (\ref{d1}) and (\ref{d2}) with $\xi_{31}=\xi_{32}=0$ to find an 
initial
stationary state of the DFFM.  It is true that the three-body loss, 
taken care of by  terms 
$\xi_{31}$ and $\xi_{32}$, is always present in the system, its effect 
is small leading to at best a small loss rate in atoms except for very 
large negative values of $a_{12}$. Hence,  for the consideration of the 
initial 
state, we could as well neglect  three-body loss. (This is why a 
Gross-Pitaevskii mean-field equation without three-body loss has been 
successfully used to study many features of a repulsive trapped 
BEC \cite{11}.)
However, in the final 
state, when $a_{12}$ is suddenly turned to a large negative value by a 
Feshbach resonance, the three-body loss has a dramatic effect 
responsible for 
a proper description of a 
collapse and explosion with  
a very rapid loss of atoms in a short interval of time (see figure 3).

In our numerical investigation we take
$l=1$ $\mu$m and consider the equal-mass fermions with the mass of
$^{40}$K corresponding to a radial frequency $\omega \approx 2\pi \times
83$ Hz. The present simplified mean-field model cannot predict the
suppression of collapse \cite{gehm,hei} of a DFFM in two hyperfine
states which is a true quantum many-body effect. Nevertheless, it leads
to a proper description of collapse dynamics of a DFFM of two distinct
atoms. The use of equal-mass fermions only keeps the algebra simple 
specially in section 2.2, but
otherwise has no effect on the general qualitative dynamics studied in 
this paper.  
In this study, the unit of time is
$1/\omega \approx 2$ ms, and unit of length $l/\sqrt 2 \approx 0.7$
$\mu$m. 
Actually, any two different fermionic atoms can be used in
experiment, a proper quantitative treatment of which will require the 
use of 
different mass factors in the dynamical equations.

\subsection{Variational Analysis}  

To understand how the stationary states of a DFFM
are formed, we employ
a variational method for the solution of  (\ref{d1}) and (\ref{d2}) in
the symmetric case $N_1=N_2\equiv N$, while
$\phi_1/x=\phi_2/x\equiv \varphi$ satisfies 
\begin{eqnarray}\label{v1}
\biggr[-i\frac{\partial
}{\partial \tau} -\frac{\partial^2}{\partial
x^2}-\frac{2}{x}\frac{\partial}{\partial
x}
+\frac{x^2}{4}
+ \mu
\left| {\varphi}\right|^{4/3}
+ g \left
| \varphi\right|^2\biggr] \varphi=0,
 \end{eqnarray}
where $g= 6 \sqrt 2 N a_{12}/l$ and $\mu = 3(3\pi N/2)^{2/3}$
\cite{varia}. Here
we have
set the absorptive terms to zero for stationary states.  
We consider  the following trial 
Gaussian
wave function for the solution of   (\ref{v1})
\cite{varia}
\begin{equation}\label{twf}
\varphi(x,t)=  A(t)\exp\left[-\frac{x^2}{2R^2(t)}
+\frac{i}{2}{ \beta(t) }x^2+i\alpha(t)
\right],
\end{equation}
where $A(t)$,  $R(t)$, $\beta(t)$, and $\alpha(t)$ are the
normalization, width, chirp, and
phase, respectively.  The normalization condition $ \int_0 ^\infty dx
x^2
\varphi^2(x,t) =1$ sets 
$A(t)=[\pi^{1/4}R^{3/2}(t)/2]^{-1}$.  The Lagrangian density for 
generating  
(\ref{v1})   is \cite{varia}
\begin{eqnarray}
{\cal L}(\varphi)&=&\frac{i}{2}\left(\dot \varphi  \varphi^*
- \dot  \varphi^*  \varphi
\right)-
\left|\frac{\partial
  \varphi}{\partial x} \right|^2  -
\frac{x^2}{4}| \varphi|^2\nonumber \\
&-& \frac{1}{2} g | \varphi|^4- \frac{3}{5} \mu |\varphi|^{10/3} ,
\end{eqnarray}
where  the
overhead dot represents time derivative.
The trial wave function (\ref{twf}) is
substituted in the Lagrangian density and the
effective Lagrangian $L_{\mbox{eff}}$
is calculated via  $L_{\mbox{eff}}= \int {\cal
L}(\varphi)
d ^3 x:$ 
\begin{eqnarray} 
L_{\mbox{eff}}=\frac{\pi^{3/2}A^2(t)R^5(t)}{2}\biggr[-\frac{3}{2}\dot
\beta(t)
-\frac{g}{2\sqrt 2} \frac{ A^2(t)}{R^2(t)}\nonumber
\\-\frac{9\sqrt 3}{25 \sqrt 5}\mu \frac{A^{4/3}(t)}{R^2(t)}
-\frac{2\dot \alpha(t)}{R^2(t)} -\frac{3}{R^4(t)}-3\beta^2(t) -
\frac{3}{4} \biggr].
\end{eqnarray}

The generalized Lagrange equations for this effective Lagrangian 
given by  \cite{gold}
 \begin{equation}
\frac{d}{d t}\frac{\partial L_{\mbox{eff}}}{\partial\dot \gamma(t)}=
\frac{\partial L_{\mbox{eff}}}{\partial \gamma(t)},
\end{equation}
with $\gamma(t)$ representing   $\alpha(t)$, $A(t), \beta(t),$ and
$R(t) $
are  written explicitly as 
\begin{equation}\label{e1}
\pi^{3/2}A^2R^3= \mbox{constant}=4\pi,
\end{equation}
\begin{eqnarray}\label{e2}
3\dot \beta+\frac{4\dot
\alpha}{R^2}+\frac{6}{R^4}+6\beta^2+\frac{3}{2}=- 
\frac{\sqrt 2gA^2}{R^2}-\frac{6\sqrt 3}{5 \sqrt 5}\frac{\mu A^{4/3}}{R^2}
,\nonumber \\ \end{eqnarray}
\begin{eqnarray}\label{e3}
\dot R= 2 R \beta,
\end{eqnarray}
\begin{eqnarray}\label{e4}
5\dot \beta+\frac{4\dot
\alpha}{R^2}+\frac{2}{R^4}+10\beta^2+\frac{10}{4}=-
\frac{gA^2}{\sqrt
2R^2}-\frac{18\sqrt 3}{25\sqrt 5}\frac{\mu A^{4/3}}{R^2},\nonumber \\
\end{eqnarray}
where the time dependence of different observable is suppressed.
Eliminating $\alpha$ between   (\ref{e2}) and (\ref{e4}) one obtains
\begin{eqnarray}\label{e5}
2\dot \beta= \frac{4}{R^4}-4\beta^2+ \frac{gA^2}{\sqrt
2R^2}+\frac{12\sqrt 3}{25\sqrt 5}\frac{\mu A^{4/3}}{R^2}
-1.
\end{eqnarray}

\begin{figure}
 
\begin{center}
\includegraphics[width=1.\linewidth]{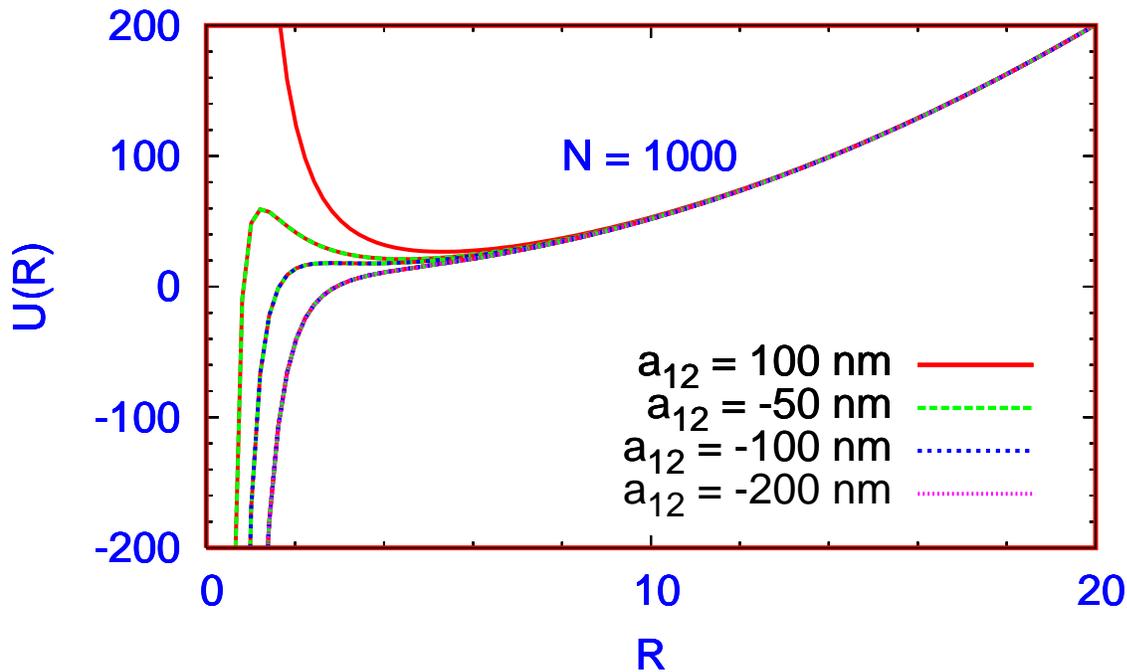}
\end{center}

\caption{ The effective potential $U(R)$ of  
(\ref{eff}) vs. $R$ for different $a_{12}$ and $N=1000$ and $l=1$ $\mu$m. 
}

\end{figure}

 From   (\ref{e3}) and  (\ref{e5}) we get the following
second-order
differential equation for the evolution of the width $R$
\begin{eqnarray}\label{el}\frac{d^2R}{dt^2}
=\frac{4}{R^3}+\frac{4 g}{\sqrt{2
\pi}R^4}+\frac{12\mu 4^{2/3}\sqrt 3}{25\pi^{1/3} \sqrt 
5R^3}
-R,
\end{eqnarray}
\begin{eqnarray}
= -\frac{d}{dR}\left[\frac{2}{R^2}+ \frac{4g}{3\sqrt{2
\pi}}\frac{1}{R^3}+ \frac{6\mu  4^{2/3}\sqrt 3}{25\pi^{1/3} \sqrt 
5R^2} 
+\frac{R^2}{2}
\right]. \label{el2}
\end{eqnarray}
The quantity in the square brackets of  (\ref{el2}) is the effective
potential  $U(R)$
of the  equation of motion:
\begin{equation}\label{eff}
U(R)=\frac{2}{R^2}+ \frac{4g}{3\sqrt{2
\pi}}\frac{1}{R^3}+ \frac{6\mu  4^{2/3}\sqrt 3}{25\pi^{1/3} \sqrt
5R^2}
+\frac{R^2}{2}. 
\end{equation}
Small oscillation of a stationary state 
around a
stable configuration is  possible when there is a minimum in this
effective
potential determined by a zero of  (\ref{el}). This condition
yields the variational width from which the variational solution for
the wave function is obtained via  (\ref{twf}).

\begin{figure}
 
\begin{center}
\includegraphics[width=1.\linewidth]{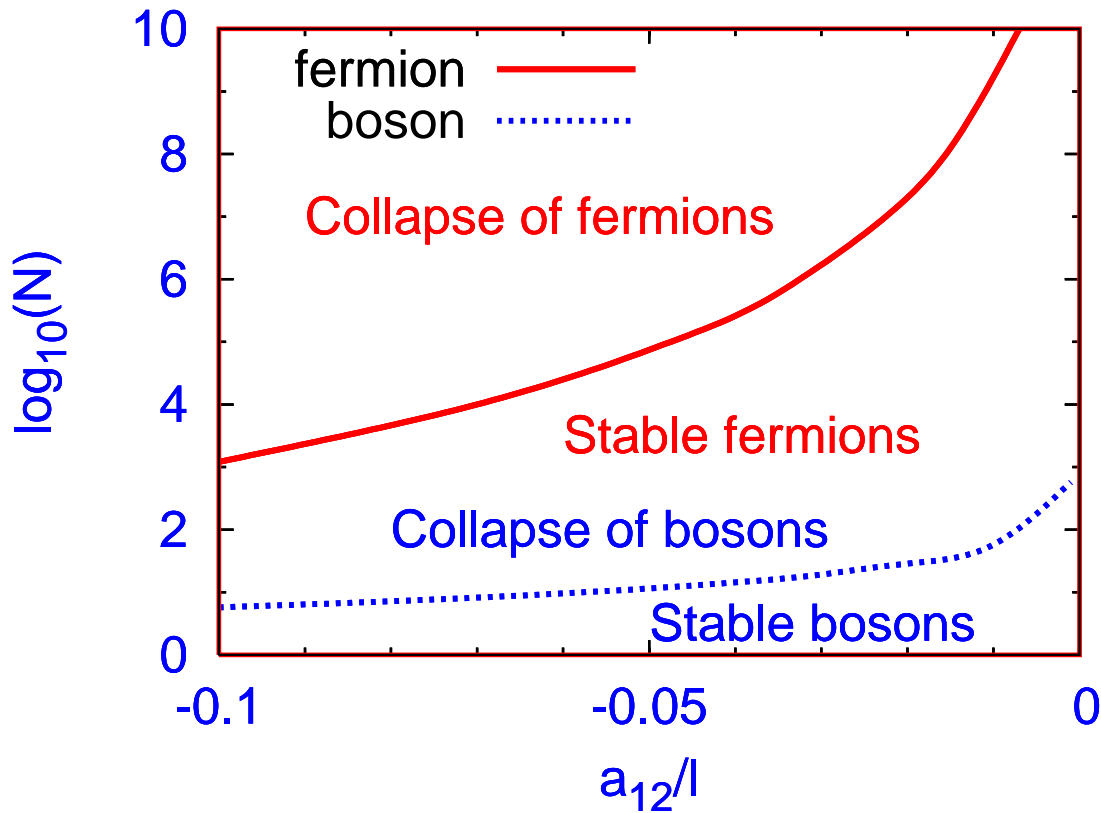}
\end{center}

\caption{The phase diagram for collapse where we plot
log$_{10}(N)$ vs. $a_{12}/l$ for fermion-fermion mixture (full
line) and bosons (dotted line). The line separates the regions where
collapse is present and  absent.}

\end{figure}

In figure  1 we plot the effective potential  $U(R)$ of 
(\ref{eff})  for   different $a_{12}$, 
$N=1000$ and $l=1$ $\mu$m.  For positive (repulsive) $a_{12}=100 $ nm,  
$U(R)$ 
leads to a confining well with a minimum at $R=R_0=5.3$, so that one 
could 
have a stable DFG  of width $R_0=5.3  $. The variational
profile for 
this 
function is 
\begin{equation} \label{var}
\varphi (x) = 
\frac{2}{\pi^{1/4}R_0^{3/2}}\exp\left[-\frac{x^2}{2R_0^2}\right].
\end{equation}
In  figure 1, as $a_{12}$ turns  gradually negative (attractive), the 
infinite wall near $R=0$ of  $U(R)$ is gradually lowered and for a 
sufficiently attractive scattering length $a_{12}\approx -100$ nm, 
this 
wall 
disappears
completely 
and one has the possibility of collapse for 
 $a_{12}<  -100$ nm. The minimum in $U(R)$ first becomes a
point 
of inflection for $a_{12}\approx -100$ nm and then disappears.  The
profile of the effective potential in  figure  1 is similar to the same in
the
bosonic case \cite{varia,11}.

Next we show  the phase diagram for collapse for $N_1=N_2=N$ using the
variational
approach in  figure  2, where  we plot log$_{10}(N)$ vs. $a_{12}/l$. 
The line
separates the plot in two regions. In the upper half of the plot collapse
is possible and in the lower half we have stable configurations.  The
phase diagram of  figure  2 is quite similar to that obtained in
Ref. \cite{phase} in a study of the stability of a DFFM.  In case of bosons in a spherically-symmetric
trap the line of stability is given by $Na/l=-0.575$ \cite{11} and is also
shown in
 figure  2. As expected, for a fixed $|a/l|$ a much larger number of 
fermions 
can be accommodated in a stable state.

\section{Numerical Results}

In this section we present results on collapse from a numerical solution
of  (\ref{d1}) and (\ref{d2}).
After some experimentation 
we take in the initial  DFFM
$N_1=1000$, $N_2=2000$, and $a_{12}= 100$ nm, so that $a_{12}/l=0.1$. 
This
corresponds to nonlinearities ${\cal N}_{11}=843$, ${\cal
N}_{22}=1338$,
${\cal N}_{12}= 200$ and ${\cal N}_{21}=
100$. The collapse dynamics is sensitive to the loss rates
$K_{31}$ and $K_{32}$.    
As these loss rates 
are not experimentally  known,  
in  the present simulation we take them to be equal: $K \equiv 
K_{31}=K_{32}$,
and 
consider  different values of $K$.

\begin{figure}
 
\begin{center}
\includegraphics[width=1.\linewidth]{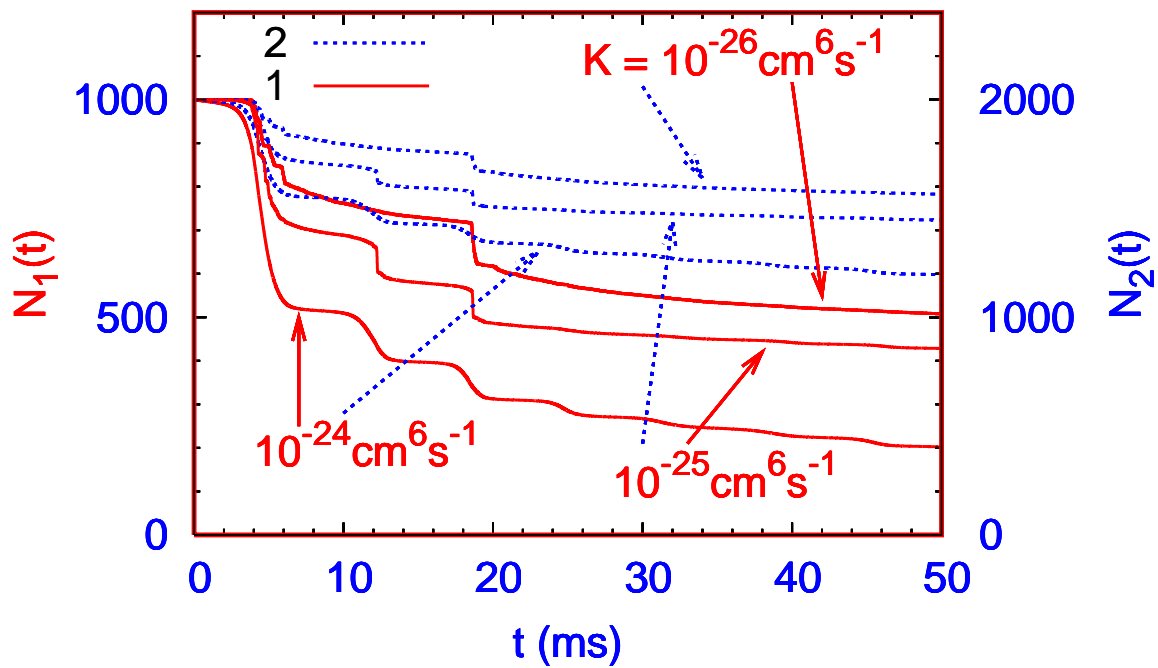}
\end{center}

\caption{ The evolution of  fermion numbers 
$N_j(t)$ of the two components  vs. time 
 during a collapse initiated by a jump in scattering
length $a_{12}$ from 100 nm to $-200$ nm  in a  DFFM
of $N_1=1000$ and
$N_2=2000$
for three-body loss rates 
$K=  10^{-26}$ cm$^6$s$^{-1}$,  $ 10^{-25}$
cm$^6$s$^{-1}$, and 
$  10^{-24}$ cm$^6$s$^{-1}.$ The dotted (blue) curves
refer to fermion 2
and the solid (red)  curves to 
fermion 1. The
curves are labeled by their respective $K$ values.  }

\end{figure}

Now we consider the collapse of fermions initiated by a sudden jump in the
fermion-fermion scattering length from $a_{12}=100$ nm to 
$-200$ nm
which
can be implemented near a fermion-fermion Feshbach resonance, 
observed in fermionic systems
 \cite{fsff}.
This
resonance should enable an experimental control of the interspecies
interaction \cite{fs} and hence can be used to increase the 
attractive
force between  interspecies   fermions by varying a background magnetic
field,  
which in turn increases the
attractive
nonlinearities $6\sqrt 2 {\cal N}_{12}$ and $6\sqrt 2 {\cal N}_{21}$ 
in
 (\ref{d1})  and (\ref{d2}). If these attractive nonlinear terms can
overcome 
the repulsive nonlinearities in these equations it is possible
to have a collapse of fermions. 

\begin{figure}
 
\begin{center}
\includegraphics[width=1.\linewidth]{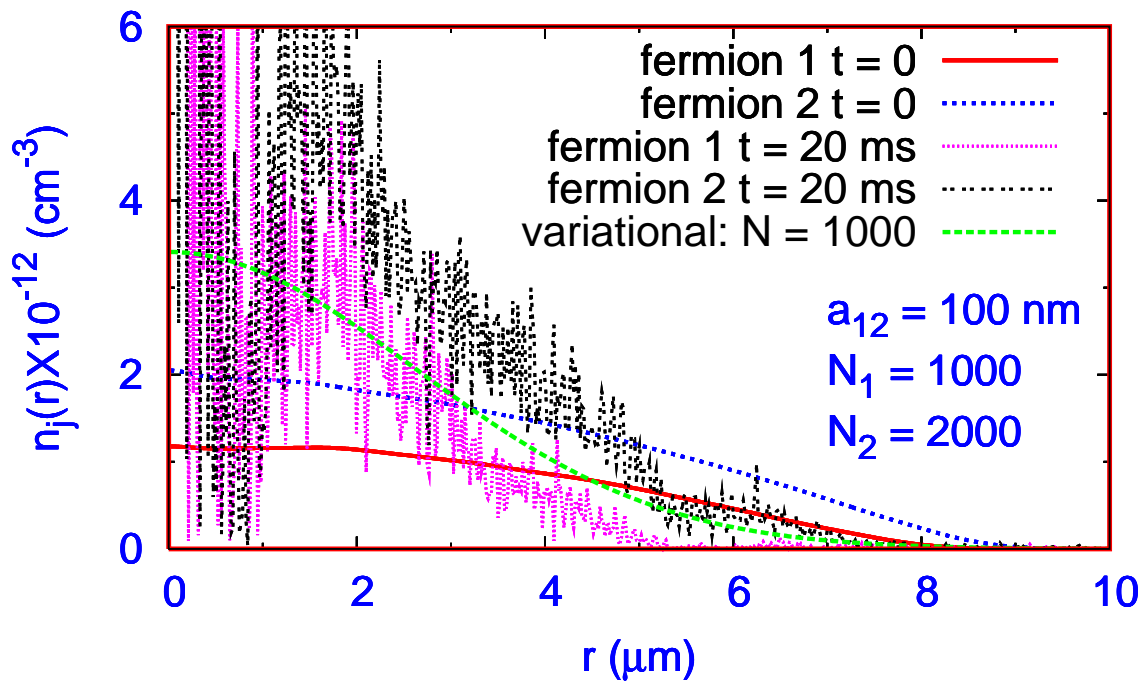}
\end{center}

\caption{ The   fermion density $n_j(r)$ at $t=0,20$
ms 
before and during the  collapse exhibited in  figure  3 
for $K=10 ^{-26}$ cm$^6$
s$^{-1}$. The density calculated from the variational profile of the wave
function (\ref{var}) for $R_0=5.3$ and $N=1000$ is also shown,}

\end{figure}

Due to the three-body loss  terms in  (\ref{d1}) and (\ref{d2}) the
number of 
fermions decay with time. When the net
nonlinear attraction
in these equations is small there is a smooth and steady decay of number
of
atoms. However, when the  net
nonlinear attraction is jumped to a large value,
the steady decay of number of
atoms develops into a violent decay called collapse. 
When this happens, the DFFM
 loses a significant fraction of atoms in a
small interval of
time (milliseconds) after which a remnant DFG  with a
reasonably
constant number of atoms is formed. Also, during and immediately
after collapse, the fermion density function becomes  unsmooth
and spiky in nature as opposed to a reasonably smooth  function in the
case of a steady decay. This also happened in the collapse of a BEC 
\cite{don}.

We study the evolution of  fermion numbers in the
DFFM 
 from time $t=0$ to $t=50$ ms after a sudden
jump in the
scattering length from $a_{12}=100$ nm to $-200$ nm at $t=0$. In agreement
with the variational analysis of last section we find that this jump in
the scattering length leads to collapse.  The
evolution of fermion numbers after the jump in scattering length $a_{12}$
depends on the value of the three-body loss
rate $K$. We study the sensitivity of the result on $K$ by
performing the calculation for different loss rates. In  figure  3 we plot 
the
evolution of the fermion numbers for loss rates: $K
= 10^{-26}$ cm$^6$s$^{-1}$, $10^{-25}$ cm$^6$s$^{-1}$, and
$10^{-24}$ cm$^6$s$^{-1}$. 
With the increase of $K$,  the decay
rate increases, although the results for different $K$ are qualitatively
similar. We see in  figure  3 that, in all cases, the 
number of  fermions decays
rapidly and attain an approximately  constant (remnant)  value  in less
than 50
ms as in the case of bosons \cite{don,th1}. We used a space step of 
0.025 in the numerical solution of 
 (\ref{d1}) and (\ref{d2}) and found that this step size was
sufficient to reach a converged result even during collapse. In our
previous study on the collapse of a BEC of $^{85}$Rb \cite{ska,th1} we
found that 
even a much larger step size of 0.1 led to converged result for the number
of atoms in the remnant
in
quantitative agreement with experiment \cite{don}. 
This gives assurance on the reliability of the present calculation.

\begin{figure}
 
\begin{center}
\includegraphics[width=1.\linewidth]{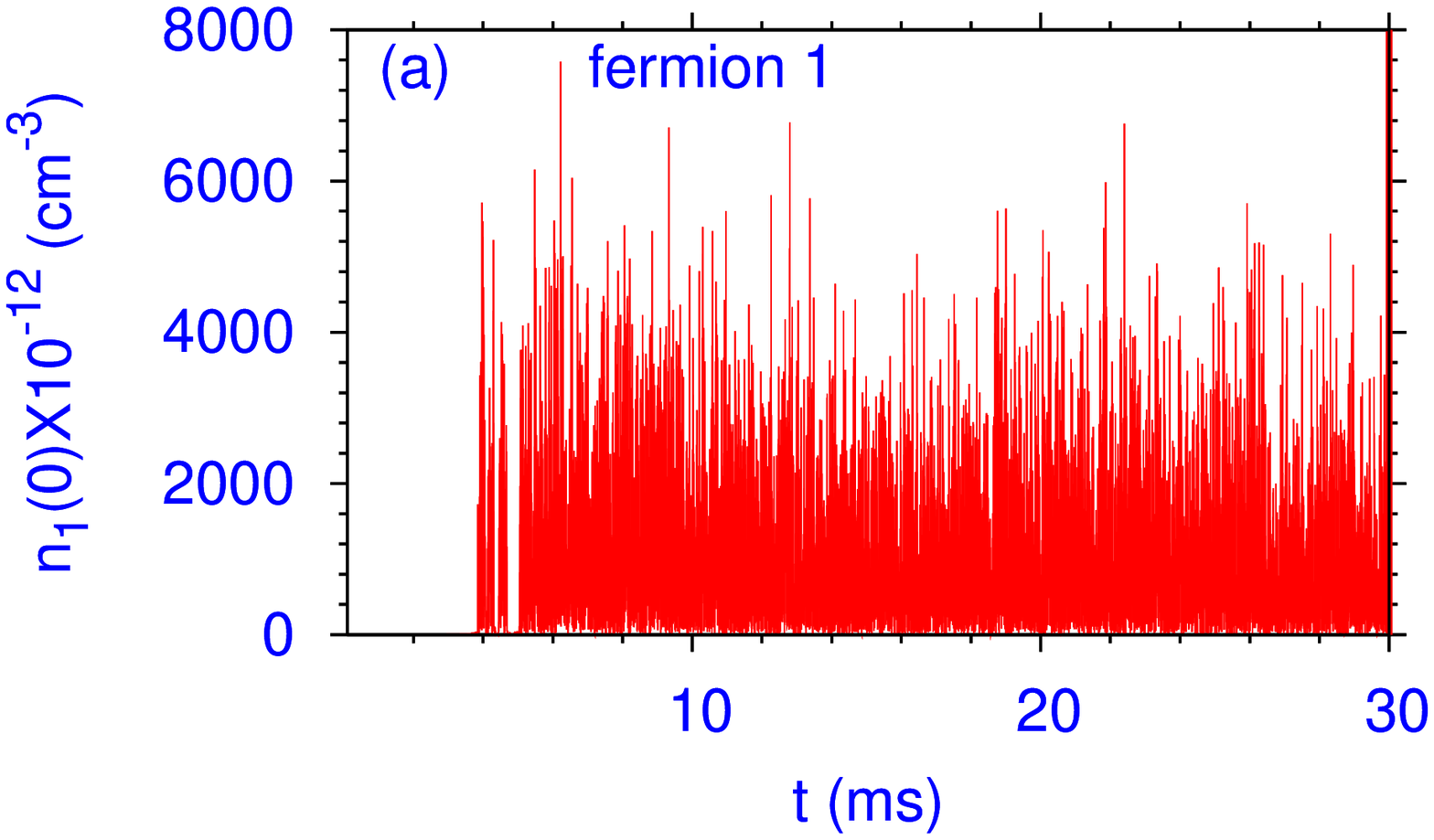}
\includegraphics[width=1.\linewidth]{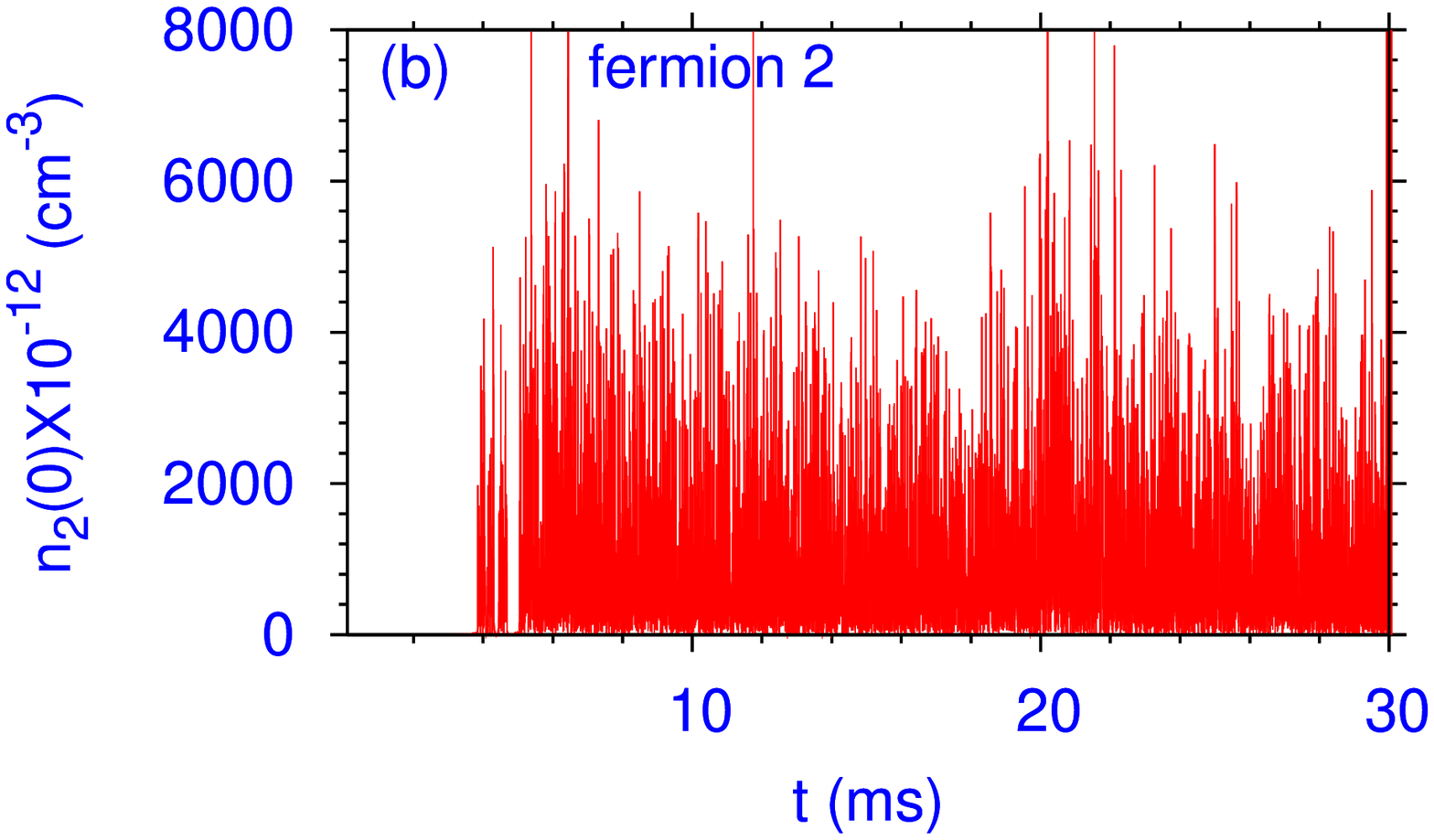}
\end{center}

\caption{Evolution of    central density $n_j(0)$
of  fermion $j=$ (a) 1 and (b) 2
 during the collapse exhibited in
 figure  3 for $K=10^{-26}$ cm$^6$s$^{-1}$.}

\end{figure}

 In  figure  4
 we plot the fermion probability densities
at times $t=0$ and $t=20$ ms.
A close look at  figure  4 reveals that before collapse at $t=0$ the
fermion densities are smooth. We have also plotted here the density
corresponding to the variational profile (\ref{var}) for $R_0=5.3$ for
fermion number $N=1000$. Although the ranges of the exact and variational
densities agree with each other, the functions do not agree 
well. This is understandable as the exact profile of the 
DFG  is very
different from the Gaussian trial function used in variational
calculation. Although the fermion densities at $t=0$ are smooth, 
the fermion densities  during and after collapse  have an entirely 
different profile. 
As expected the  densities  are highly peaked in the central ($r=0$) 
region and develop spikes. Near $r=0$ the densities could be two to three
orders of magnitude larger than those for larger $r$ values (see  figure  
5).
However, they extend over a large distance
too. The final spiky function indicates the collapse, in contrast to a
smooth final  function corresponding to a  steady loss of
atoms. The collapse is a quick process lasting at most a few tens of
milliseconds when a significant fraction of atoms are lost. For example,
in  figure  3 for $K=10^{-24}$ cm$^6$s$^{-1}$, the collapse lasts for the
first 
25 ms when most of the atoms are lost. After this 
interval the rate of loss of atoms is reduced and 
remnant a DFG  
with a roughly constant number of atoms are formed.

To confirm further the  collapse in  figures 3 and 4 for $K=
10^{-26}$
cm$^6$s$^{-1}$, we plot 
in  figure  5  the evolution of the central probability  density
$n_j(0)$
of fermion $j$ during collapse. 
We note a very strong fluctuation of a very large  central density
reminiscent of
collapse in both  components. The central density is three orders of
magnitude larger than the equilibrium density in  figure  4. 
Similar fluctuations 
were noted in the  collapse of a pure BEC 
\cite{th2}  as well as a DBFM
\cite{ska}. Such a
strong fluctuation of the central density  could not be due to a weak
evaporation of the  DFFM  due to recombination.

\begin{figure}
 
\begin{center}
\includegraphics[width=1.\linewidth]{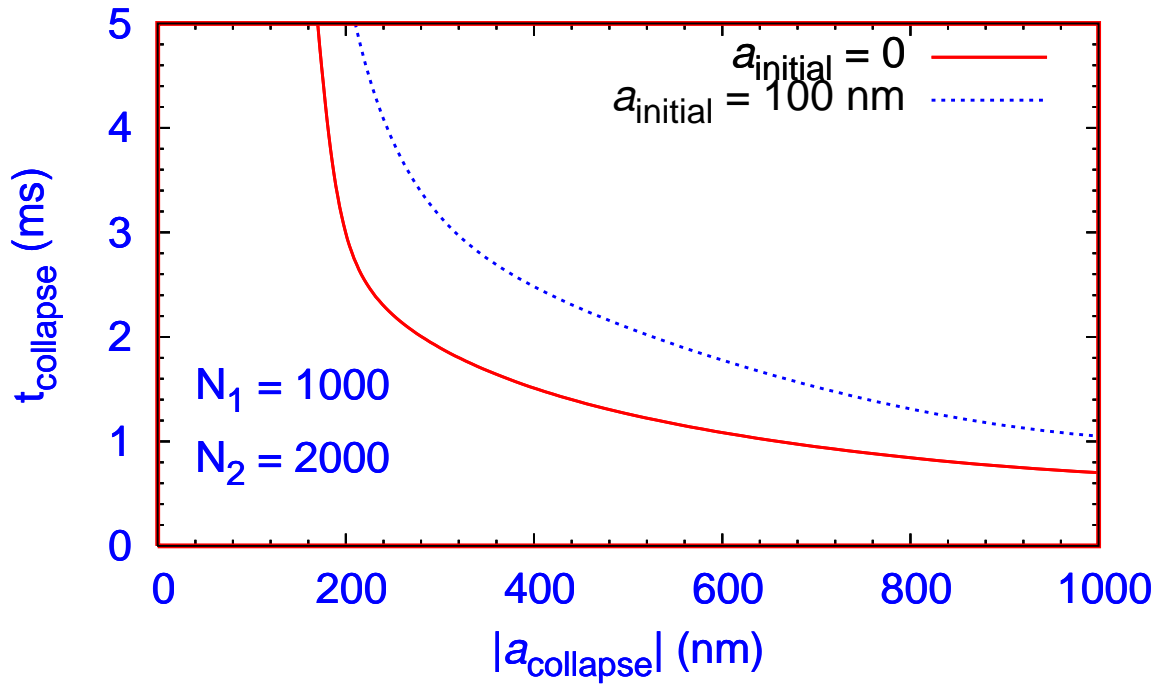}
\end{center}

\caption{ The evolution of time to collapse 
$t_{\mbox{collapse}}$ vs. final scattering length $a_{\mbox{collapse}}$
for $a_{\mbox{initial}} = 0$ and 100 nm for a  DFFM
with $N_1=1000$, 
$N_2=2000$ and 
$K=  10^{-26}$ cm$^6$s$^{-1}$.   The
curves are labeled by their respective $a_{\mbox{initial}}$ values.}

\end{figure}

From  figure  3  we find that the number of fermions remains practically 
constant during the first 4 ms or so after jumping the scattering 
length from 100 nm to $-200$ nm
indicating that the collapse starts only 
after this interval of time.  
This is confirmed from the plot of central densities in  figure  5 where
we 
see that very large values of central density also appear after 
an  
interval of time called ``time to collapse". A similar phenomenon was also 
observed in the collapse 
of bosons \cite{don}. Next we study an evolution of this time to 
collapse ($t_{\mbox{collapse}}$) with changing initial 
($a_{\mbox{initial}}$) and final ($a_{\mbox{collapse}}$) scattering 
lengths. This is shown in  figure  6 for two values of
$a_{\mbox{initial}}$, 
where we plot $t_{\mbox{collapse}}$ vs. $a_{\mbox{collapse}}$ for 
$N_1=1000$,  $N_2=2000$ and $K=10^{-26}$ cm$^6$s$^{-1}$. 
The time to collapse is large for a small jump in the scattering length
and reduces when the jump in the scattering length is increased, as also
observed in the
case of bosons \cite{don}.

\begin{figure}
 
\begin{center}
\includegraphics[width=1.\linewidth]{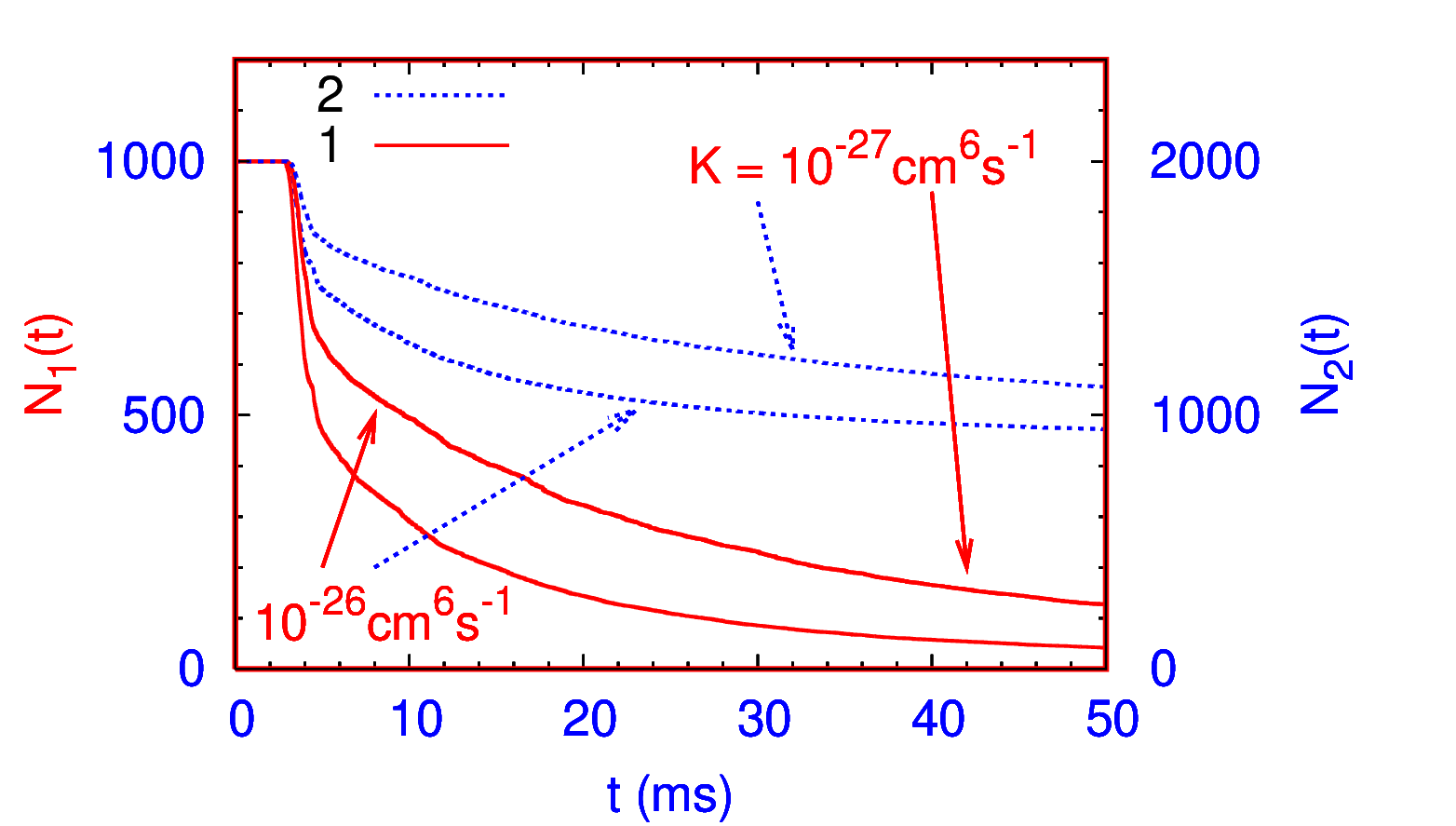}
\end{center}

\caption{ The evolution of fermion numbers 
$N_j(t)$
 during collapse initiated by a jump in  scattering
length $a_{12}$ from 100 nm to $-300$ nm  for
$K=  10^{-27}$ cm$^6$s$^{-1}$,  $ 10^{-26}$
cm$^6$s$^{-1}$, and
  $10^{-25}$ cm$^6$s$^{-1}.$ 
 The dotted (blue) curves
refer to fermion 2
and the solid (red)  curves to
fermion 1. The
curves are labeled by their respective $K$ values.}

\end{figure}

One interesting aspect of  figure  3 is the appearance of a revival of 
collapse. The fermion number after the primary collapse remains
approximately constant 
for an interval of time and then again reduces abruptly. This revival
of collapse takes place several times. A similar revival of 
collapse was noted in the fermion component in a
numerical simulation in a DBFM \cite{ska}
and was confirmed later in an experiment \cite{bongs}
on the
$^{87}$Rb-$^{40}$K DBFM.
To study the  revival of
collapse further
in a  DFFM we considered  a different
jump in the scattering length. 
For the same initial state of  figure  3 we now
consider a jump in $a_{12}$ from $100$ nm to $-300$ nm and the dynamics
is reported in  figure  7 for different $K$ values. We find that the
revival of collapse has practically disappeared  in this case. If the 
collapse is initiated by a  
small jump  in the scattering length, the initial collapse is less 
violent. However, after this initial milder collapse the  
DFFM cannot 
reach an equilibrium state and it remains large and attractive. 
Consequently, the  DFFM undergoes further
collapse(s). On the other hand, 
if the collapse is initiated by a large jump in the scattering length, 
the initial collapse is very violent  through which  the  
DFFM
 gets 
rid of a very large number of atoms. Consequently, the  DFFM reaches  
reasonably small and cold remnant states  which do not further undergo 
collapse and one has
one primary collapse.  
This is clear from  figures  3 and 7. In  figure  3 after the first 
collapse the DFFM loses a smaller percentage of atoms whereas in
 figure  7 a large 
percentage of atoms are lost after the primary collapse.

\begin{figure}
 
\begin{center}
\includegraphics[width=1.\linewidth]{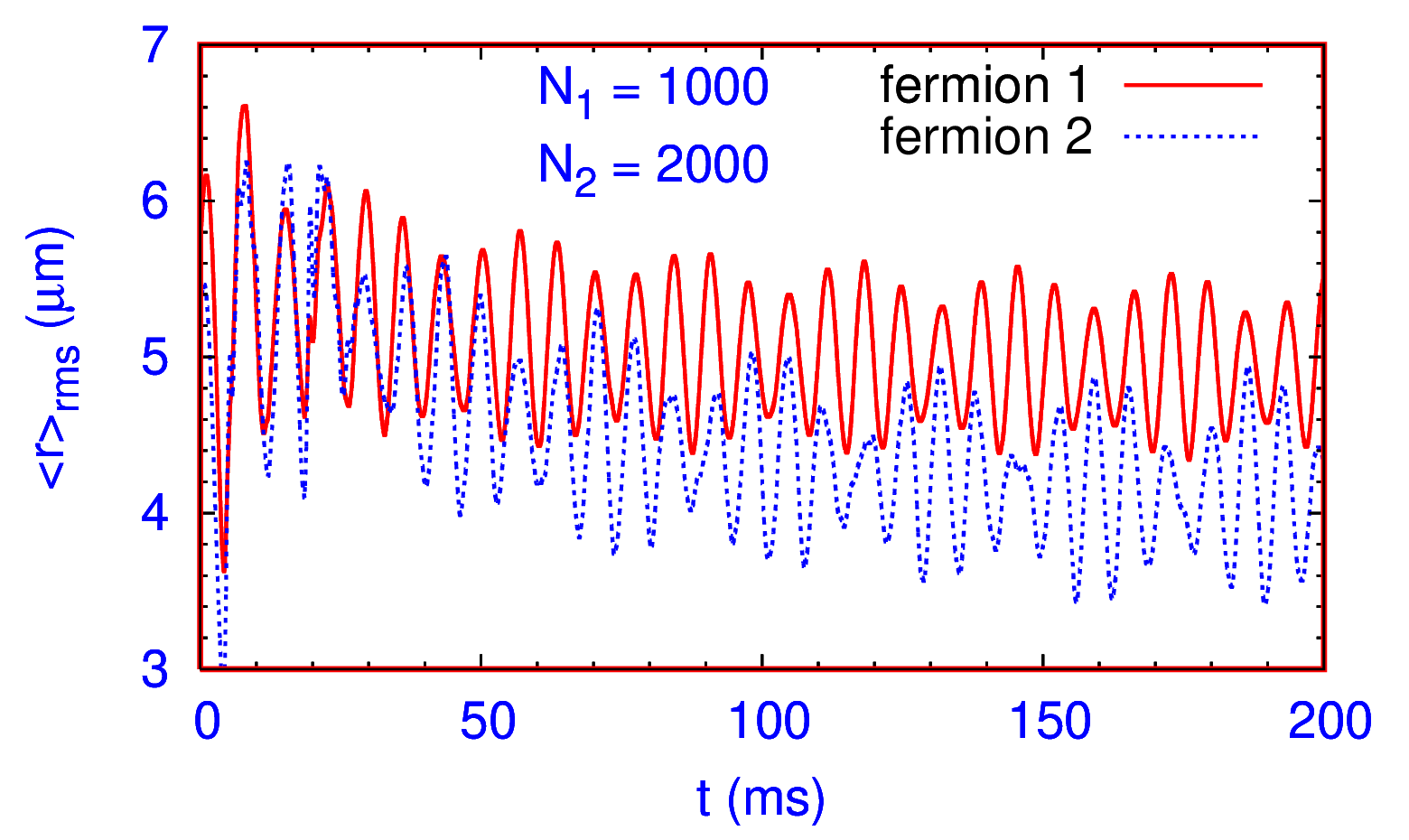}
\end{center}

\caption{ The rms radii of the two components  vs. 
time before and 
during collapse exhibited in  figure  3  for $N_1=1000$ and $N_2=2000$ 
initiated by a jump in  scattering
length $a_{12}$ from 100 nm to $-200$ nm  for
$K=  10^{-26}$ cm$^6$s$^{-1}$. 
 The dotted (blue) curves
refer to fermion 2
and the solid (red)  curves to
fermion 1. }

\end{figure}

It was found in the experiment on collapse on a BEC \cite{don} that
during collapse the root mean square (rms) sizes of the condensate
execute periodic oscillation with approximately 
twice the frequency of the trap. A breathing  oscillation of same
frequency was found in a BEC when a small perturbation was applied
\cite{sk1}.
In
dimensionless unit, the angular frequency of the trap is $\omega = 1$,
corresponding to a (linear) frequency of $1/(2\pi)$.  The 
observed frequency of
oscillation of rms size was $1/\pi$ \cite{don} $-$ twice the trap 
frequency. In actual time unit the frequency of oscillation of rms sizes
during collapse  corresponds to 
$1/(2\pi)$ ms$^{-1}$  $\approx 0.16$ ms$^{-1}$. 
We also investigated if such
oscillation existed in the present case in the  
DFFM during the 
collapse. 
In  figure  8 we plot the rms
radii of the two  components of the collapsing  DFFM
and find that they also execute
quasi-periodic oscillation. The calculated frequency from  figure 8 is
0.145
ms$^{-1}$ close to that found in the case of bosons, e.g.  0.16
ms$^{-1}$. The difference could be due to the coupled nature of the
hydrodynamic 
equations as well as the very large nonlinearity for fermions.

\section{Summary}
 
We suggested a coupled set of time-dependent hydrodynamic
equations for a trapped  DFFM including the effect of three-body 
recombination.
The present time-dependent formulation permits us to study the
non-equilibrium dynamics of a  DFFM. Using a
variational analysis as well
as a numerical solution of our model, 
we study, for an attractive
inter-species fermion-fermion interaction, the collapse 
in a DFFM composed of two types of nonidentical atoms.
The collapse of a  DFFM of two different atoms
 can be realized
experimentally 
by jumping the inter-species
scattering length  to a large negative value by exploiting a
fermion-fermion Feshbach resonance \cite{fsff}.  The collapse dynamics is
strongly
dependent
on the three-body loss rate $K$ and we present results for different loss 
rates. We note the possibility of  a revival of collapse in a  
DFFM
as in a previous simulation \cite{ska} on a 
DBFM,
confirmed later in an  
experiment \cite{bongs} on
$^{87}$Rb-$^{40}$K.  We find that a revival of collapse in a 
DFFM takes 
place for a moderate jump in the interspecies scattering length which
disappears for a larger jump. We also study the quasi-periodic 
oscillation of
the  DFFM
with approximately twice the trap
frequency
during collapse and explosion. 

A proper treatment of a  DFFM
should be performed
using a fully
antisymmetrized many-body Slater determinant wave function \cite{yyy1} as
in the case of atomic and molecular scattering involving many electrons
\cite{ps}. However, in view of the success of the hydrodynamic model in a
description of a collapse \cite{ska}, the formation of bright \cite{fbs2}
and dark \cite{fds} solitons, and vortex states \cite{Jezek} in a
DBFM,
we
do not believe that the present study on the collapse in a  DFFM
to be so
peculiar as to have no general validity. The present study on
collapse in a  DFFM of nonidentical atoms
can be verified in future
experiments, which can really
validate the present hydrodynamic model and the related numerical study.

 
\ack

The work is supported in part by the CNPq and FAPESP  
of Brazil.

\end{document}